\documentclass[longbibliography,groupedaddress,twocolumn]{revtex4-2}

\usepackage{graphicx}
\usepackage[percent]{overpic}
\usepackage{braket}

\usepackage{multirow}
\usepackage{booktabs}
\usepackage[table,xcdraw]{xcolor}
\usepackage{bm}
\usepackage{blindtext}
\usepackage{subfigure}
\usepackage{commath}
\usepackage{amsmath}
\usepackage{verbatim}
\usepackage{pifont}

\usepackage[overlay,absolute]{textpos}

\newcommand\PlaceText[3]{
	\begin{textblock*}{10in}(#1,#2)
		#3
	\end{textblock*}
}

\usepackage{floatrow}
\newfloatcommand{capbtabbox}{table}[][\FBwidth]

\begin{document}


\newcommand{\myTitle}{Self-tuning transmitter for quantum key distribution using machine intelligence

} 

\author{
    Y.~S.~Lo$^{1, 2,*}$\email[]{\myEmail},
	R.~I.~Woodward$^{1}$,
	T.~Roger$^1$,
	V. Lovic$^{1, 3}$,
	T.~K.~Para\"{i}so$^1$,
	I.~De Marco$^1$,
	Z.~L.~Yuan$^{1}$,
	A.~J.~Shields$^1$}

\newcommand{\myAffiliations}{
	$^1$Toshiba Europe Ltd, Cambridge, UK\\
	$^2$Quantum Science \& Technology Institute, University College London, UK\\
	$^3$Department of Physics, Imperial College London, UK \\
}

\newcommand{\myEmail}{yuen.lo.18@ucl.ac.uk}

\title{\myTitle}

\affiliation{\myAffiliations}
\email[]{\myEmail}

\begin{abstract}

The development and performance of quantum technologies heavily relies on the properties of the quantum states which often require careful optimisation of the driving conditions of all underlying components. In quantum key distribution (QKD), optical injection locking (OIL) of pulsed lasers has recently been shown as a promising technique to realise high-speed quantum transmitters with efficient system design. However, due to the complex underlying laser dynamics, tuning such laser system is both a challenging and time-consuming task. Here, we experimentally demonstrate an OIL-based QKD transmitter that can be automatically tuned to its optimum operating state by employing a genetic algorithm. Starting with minimal knowledge of the laser operating parameters, the phase coherence and the quantum bit error rate of the system have been optimised autonomously to a level matching the state of the art.  
\end{abstract}

\maketitle

\PlaceText{12mm}{8mm}{Phys. Rev. Applied \textbf{18}, 034087 (2022); https://doi.org/10.1103/PhysRevApplied.18.034087}

\section{Introduction}


Quantum key distribution (QKD) allows two remote users to communicate with unconditional security, where no computational assumptions are imposed on potential eavesdroppers \cite{Gisin2002, bb84}. Such technology is of particular interest as conventional cryptographic schemes relying on computational complexity are increasingly becoming vulnerable due to the rapid advances in quantum computation \cite{Arute2019}. Over the past decades, QKD has moved far beyond proof-of-principle experiments, extending to quantum networks \cite{Chen2021, Dynes2019, Sasaki2011} and satellite-based QKD \cite{Yin2020, Liao2017}. Indeed, commercial QKD systems are currently available on the market and ready for real-world applications \cite{usecase}.

As the adoption of QKD technology continues to grow, so too do the demands for more robust and reliable systems. One of the key components of a QKD system is the transmitter where the quantum states are prepared. Optical injection locking (OIL) with gain-switched laser diodes has emerged as a promising technique to realise high-speed, robust and cost-effective quantum transmitters. Not only does the OIL technique improve the laser characteristics, such as a reduction in pulse timing jitter, chirping suppression and modulation bandwidth enhancement \cite{Lau2009,Liu2020, Paraiso2021}, it also enables direct phase encoding, where the phase information can be directly encoded by varying the electrical waveform applied to the lasers, thereby removing the need for conventional bulky and costly LiNbO$_{3}$ modulators \cite{Yuan2016}. OIL has been widely applied to many QKD protocols, including BB84 \cite{Roberts2018a}, coherent-one-way QKD \cite{Roberts2017}, measurement-device-independent (MDI) QKD \cite{Woodward2021a, Wei2020} and Twin-field (TF) QKD \cite{Liu2021}. Chip-based QKD exploiting OIL has also been realised \cite{Paraiso2019}, as well as OIL-based QKD transmitter designs capable of adapting to different protocols and clock rates \cite{DeMarco21}.

While OIL offers many attractive features, the underlying laser dynamics are in fact very complex \cite{Lau2009} and involve the interplay between multiple control parameters. In order to achieve stable locking condition for low-noise and high-coherence outputs, one unavoidably has to optimise a number of parameters simultaneously. Furthermore, even with the same model of laser, every laser has slightly different properties, arising from natural variations and component tolerances during manufacturing. Therefore, very often the optimum parameters determined for one system cannot be directly applied to another system, necessitating the need to optimise each system individually. It is therefore highly desirable to devise an efficient approach to tune the systems automatically and reliably without relying on specialised personnel.

The most straightforward approach would be to implement linear parameter sweeps to search for the optimum operating regime. However, the sheer size of parameter space spanned by the multiple laser parameters makes this approach infeasible. Moreover, the problem is further complicated by the complex landscape of the search space which typically exhibits multiple local optima. 

Machine learning (ML) algorithms are widely used across the fields of science and technology for various applications, including pattern recognitions and complex system controls.
In the field of QKD, ML techniques have recently been applied to estimate the optimal parameters for QKD protocols (e.g. flux intensities and sending probabilities) \cite{Wang2019,Ding2020,Lu2019} and the phase drift in an interferometer \cite{Liu2019}. In practice however, the challenges usually lie in tuning the hardware components for optimum operation, hence, a question that arises naturally is whether ML techniques can be used to interact directly with the QKD hardwares and learn to operate a QKD system in a fully autonomous way. While a wide range of avenues are available to implement machine intelligence, genetic algorithms (GAs) have been shown to be a promising candidate to achieve such goal \cite{Genty2021}. A GA is a heuristic search algorithm inspired from Darwin's theory of natural selection, which mimics the process of biological evolution to determine the `fittest individual' in performing a given task \cite{MITpress}. GAs are well known for their ability to efficiently search through a vast parameter space and locate the global optimum. The application of GAs have been demonstrated in various contexts in photonics, such as achieving stable mode-locking in fibre lasers \cite{smartlaser}, supercontinuum generation \cite{Wetzel2018} and the design of optical components \cite{Prudenzano2007, Martin1995, Kerrinckx2004}. 

Here, we experimentally demonstrate a self-tuning QKD transmitters based on a GA. This represents the first autonomous optimisation of OIL laser system using machine intelligence. To demonstrate the flexibility of the GA, our method is applied to optimise the interpulse phase coherence and the QBER for BB84 protocol. We show that performance comparable to the state of the art QKD is achieved by our self-tuning technique.

\begin{figure*}
    \centering
    \includegraphics[width=330pt]{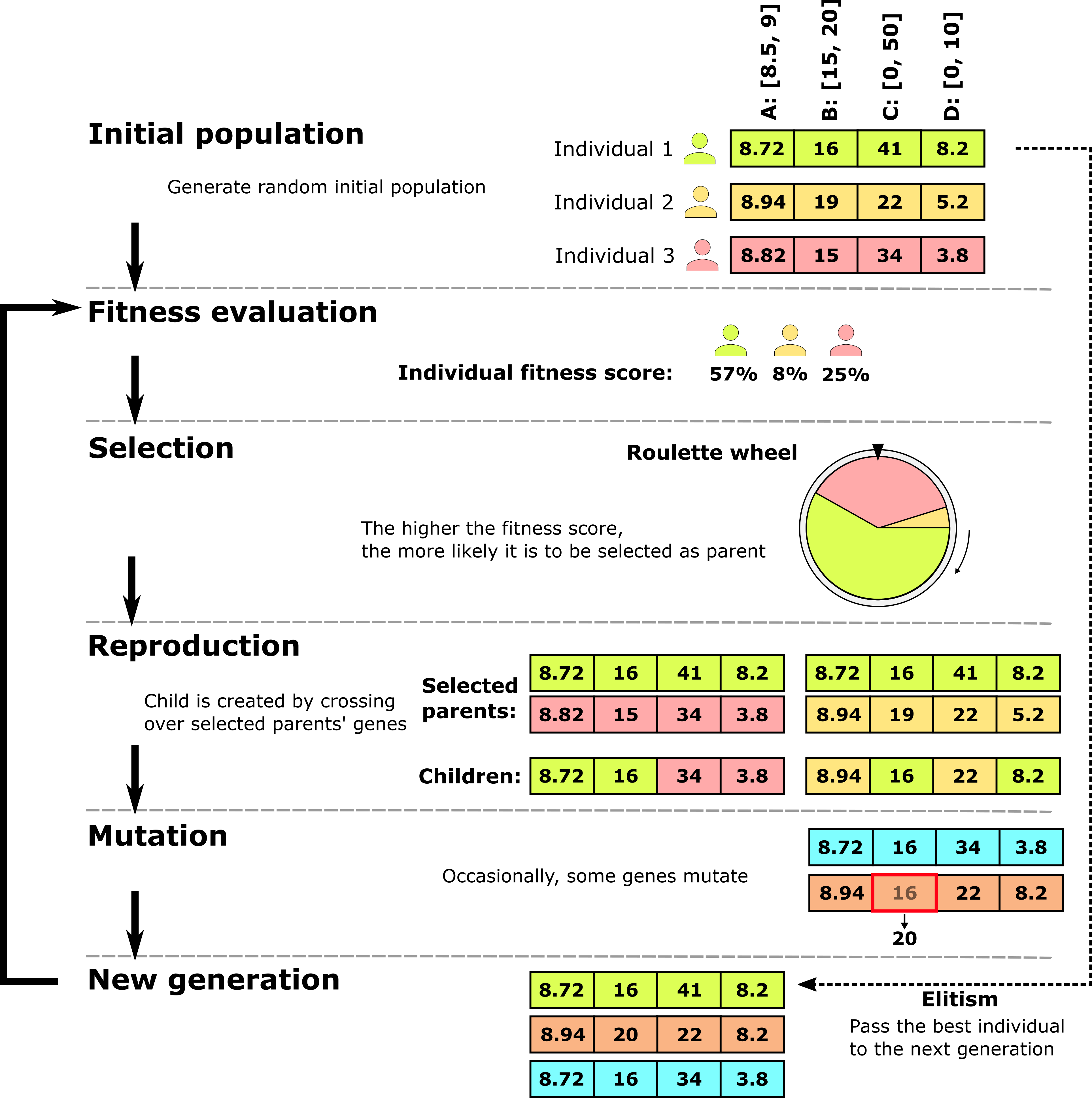}
    \caption{Schematic illustration of genetic algorithm for laser parameters optimisation.}
    \label{GA}
\end{figure*}

\section{Results}

\subsection{Genetic algorithm for parameter search}

\begin{figure*}
    \centering
    \includegraphics[width=376pt]{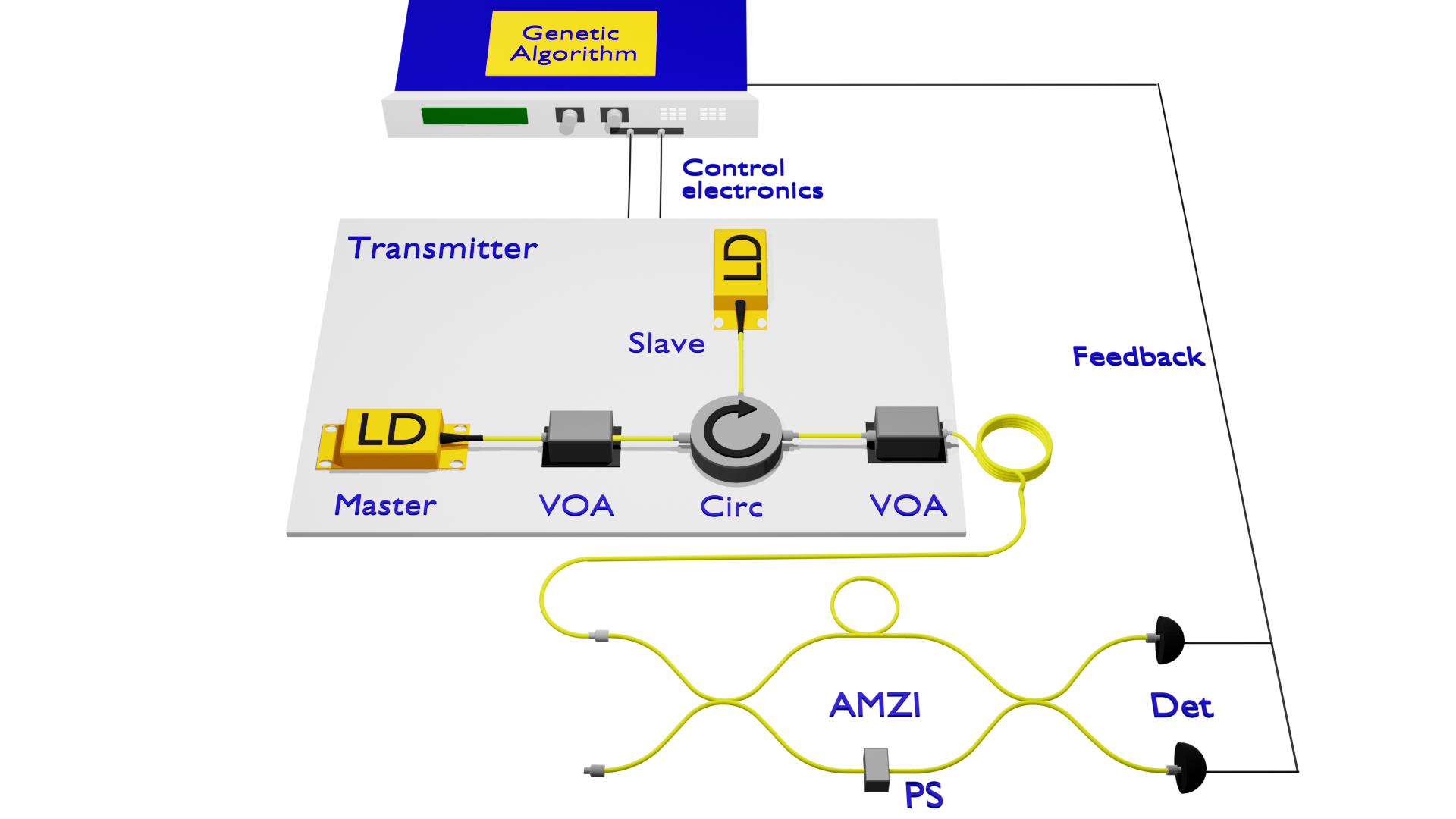}
    \caption{Experimental setup for self-tuning QKD transmitter. VOA variable optical attenuator, Circ circulator, AMZI asymmetric Mach-Zehnder interferometer, PS phase shifter, Det detector.}
    \label{setup}
\end{figure*}

The core concept of a GA is illustrated in Fig. \ref{GA}. In a GA, each possible solution is represented by an `individual'. Each individual has a set of `genes' which corresponds to the values for each parameters we aim to optimise (e.g. parameter A, B, C and D in Fig. \ref{GA}). The goal is to determine the `fittest' individual that gives the best performance. 

At the beginning, a population is initialised by randomly assigning genes to the individuals. The purpose of this step is to distribute the individuals in the search space as uniformly as possible so that any promising regions can be identified quickly as they evolve \cite{Vie2020}. Then, the fitness of each individual in the generation is evaluated by a fitness function defined by our objective. In practice, this is done by electronically setting the system parameters according to the genes of the individual and a score is assigned based on the quality of the output. In nature, natural selection favours individuals with traits that lead to more successful reproduction. Similarly in a GA, individuals are selected as parents with a probability based on their fitness score\----the higher the score, the higher the chance to be selected as parent. A child is then produced by randomly crossing over the genes from the two parents. Having sufficient genetic diversity prevents convergence towards local optima, which can be achieved via mutation where the children's genes are randomly altered with a certain probability. The crossover rate for reproduction and mutation rate are set to the typical values of 50\% and 30\%, respectively. Over successive generations, the population evolves by inheriting good genes and eliminating bad genes, until it converges to an optimum state. 

In order to speed up the convergence, we apply the concept of `elitism' where the fittest individual in the population (elite) is cloned to the next generation \cite{MITpress}. We also introduce a feature on mutation: when mutation occurs, usually the mutated gene is altered to a random values, however, here there is a 30\% chance that the mutated gene will be altered to a value close to the corresponding gene of the elite in the generation. This provides us with an additional degree of freedom to control the exploration and exploitation of the search space.

\subsection{Experimental realisation} \label{exp}

\begin{figure}
    \centering
    \includegraphics[width=236pt]{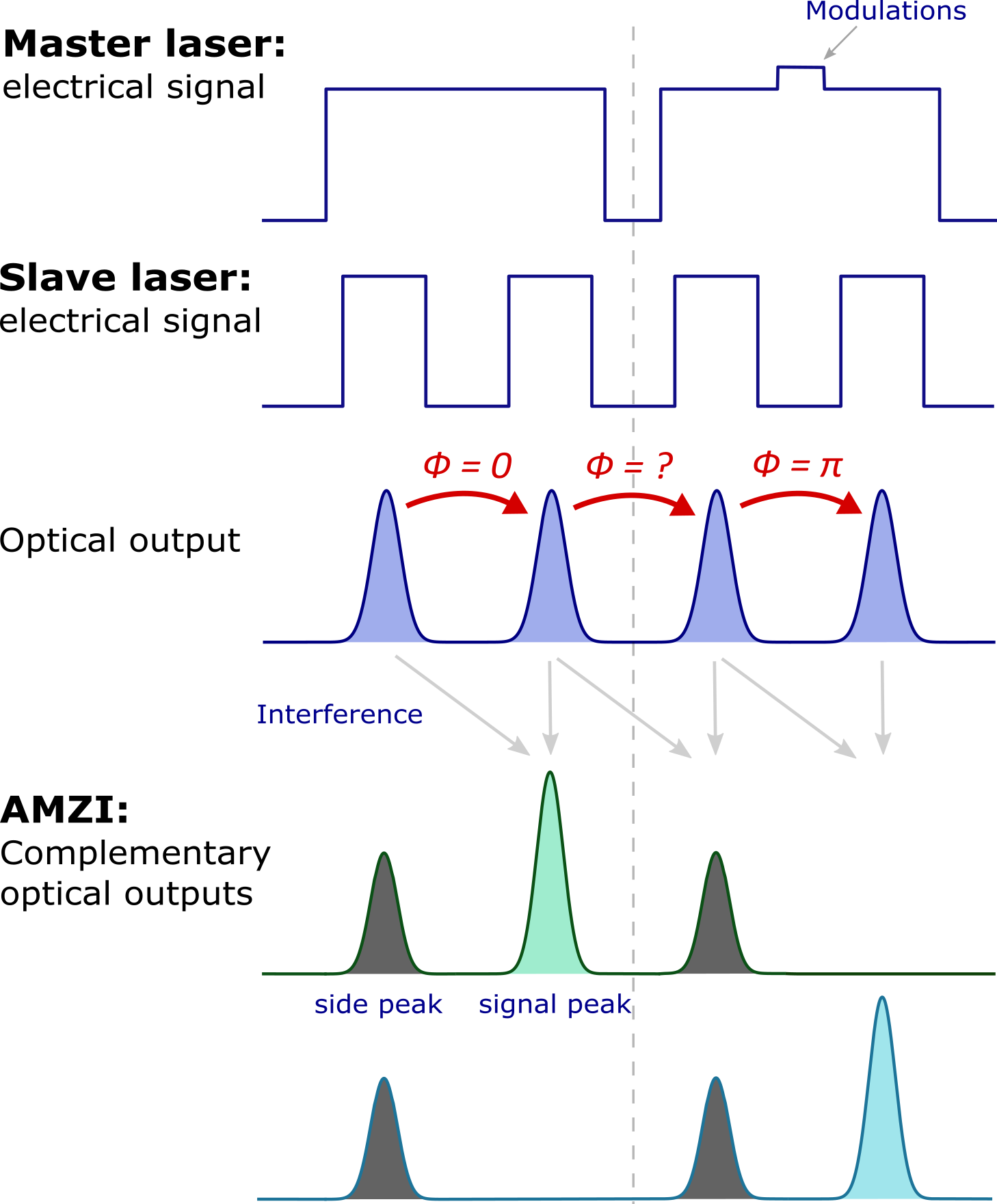}
    \caption{Principle of operation of the direct phase encoding scheme. By adding an amplitude modulation on the master laser's electrical driving signal, the phase between consecutive slave laser pulses can be implemented. The phase-encoded pulse train is decoded by an AMZI, where the phase modulation is converted to amplitude modulation suitable for direct detection.}
    \label{waveform}
\end{figure}

Our experimental setup is shown in Fig. \ref{setup}. The transmitter comprises two distributed feedback (DFB) lasers in an OIL configuration, where light from a master laser is injected into the cavity of a slave laser via an optical circulator. A variable optical attenuator (VOA) is used to control the injection power. Each laser wavelength is temperature-stabilised with an integrated thermoelectric cooler which can be tuned via a controller. RF signals and DC biases from current sources are combined using bias-tees to drive the two lasers. The master laser is gain-switched to produce a pulse train at 1 GHz. Between pulsing, the master laser is driven below the lasing threshold to ensure that each generated pulse has a random phase. The master laser pulses are injected into the slave laser, which is gain-switched at 2 GHz, generating short pulses with $\sim$70 ps duration. The RF signals of the two lasers are shown in Fig. \ref{waveform}. The two lasers are temporally aligned such that each master pulse seeds two slave laser pulses, forming the early and late time bins of a single clock cycle (i.e. a single qubit) which share the same globally random phase. 

To encode a relative phase between the two slave laser pulses, the RF signal of the master laser is modulated by adding a small amplitude perturbation during the time interval between the slave laser pulses. The perturbation changes the carrier density of the master laser cavity which in turn changes its emission frequency \cite{Bennett1990} and its phase evolution. As the slave laser pulses are seeded by the injected master photons, they inherit the phase of the master pulse. The induced phase difference in the master laser pulse is subsequently transferred onto the phase between consecutive slave laser pulses, thereby realising direct phase encoding \cite{Yuan2016}. The applied phase shift can be precisely controlled by changing the amplitude of the electrical perturbation signal. A VOA is used to attenuate the pulses before transmitting into the quantum channel. 

In the receiver, an asymmetric Mach-Zehnder interferometer (AMZI) is used to decode the relative phase between the slave laser pulses. The long arm of the AMZI has a delay of 500 ps that matches with the temporal separation of consecutive slave laser pulses. A phase shifter (PS) is used to compensate the phase drift between the two arms. Consecutive slave laser pulses can interfere constructively or destructively depending on their relative phase, thus allowing us to assign bit `0' and `1' to the two output ports. The AMZI outputs are measured with a photodiode or single photon detectors. The GA is able to control all of the laser electronics and set the values for each parameter remotely. The output of each parameter set is then measured and acts as feedback to the algorithm for evaluation. We note that the computational cost of implementing the genetic algorithm is low and a typical office-grade computer is used here, with an 8th-generation Intel Core i7 processor.

The laser parameters optimised by the GA are listed in Table \ref{parameters}. In general, to achieve stable OIL, the injection power from the master laser and the frequency detuning between the master laser and the free-running slave laser, which depend on the temperatures as well as the bias currents, need to be carefully chosen \cite{Lau2009, Liu2020}. The dynamics of OIL are more complex under gain-switching operation. It is necessary for the injection power from the master laser to be strong enough in order to overcome the influence of spontaneous emission noise on the phase in the slave laser; however, excessive injection light may create undesirable parasitic effects and degrade the performance \cite{Shakhovoy2021}. In addition, the bias current of the master laser should be set at a level that allows the laser to be driven below the threshold between each pulse for phase randomisation, meanwhile, it also affects other crucial output properties such as the phase and duration of the pulses. To transfer the phase, the two lasers need to be temporally aligned and the duration of the master laser pulse should be long enough to seed the generation of two consecutive slave laser pulses. When phase modulation is considered, the implemented phase depends on the amplitude modulation applied on master laser's driving signal. It is therefore necessary to tune all of these parameters in order to harness the benefits of OIL.

\begin{figure*}
\begin{minipage}{.84\textwidth}
    \centering
    
    \begin{overpic}[tics=10]{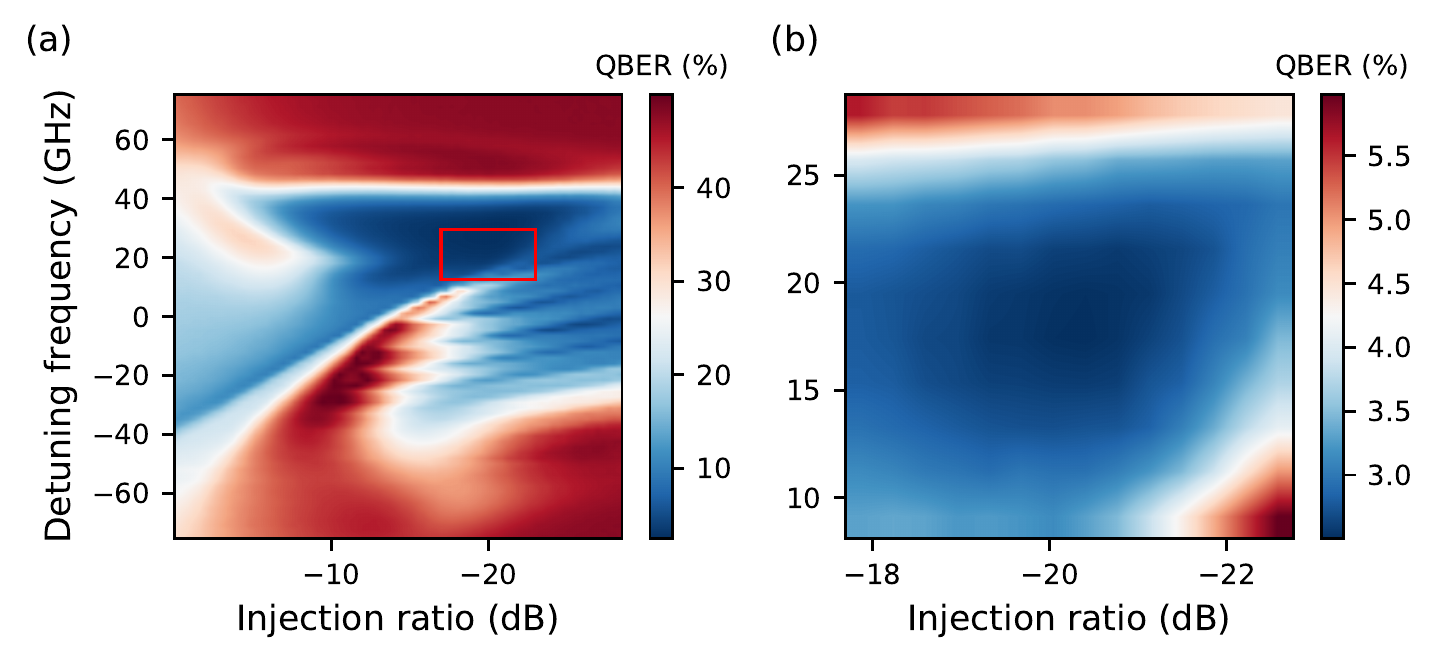} 
    
    \end{overpic}
    
    \vspace*{3mm} 
    \caption{(a) Map of experimentally measured QBER of BB84 QKD protocol as a function of laser detuning frequency and injection ratio. The red box indicates the promising region for optimum operation. (b) Zoomed-in plot of the enclosed region.}
    \label{scan}
    \end{minipage}
    
\end{figure*}

To investigate the complexity of the laser dynamics, we experimentally measure the QBER for BB84 QKD protocol as a function of the frequency detuning between the two lasers and the injection ratio (defined as the ratio between injected master power and free-running slave power), with all other parameters fixed at pre-determined optimum values, as shown in Fig. \ref{scan}. The promising operating region is indicated with a red box in Fig. \ref{scan}a and further enlarged in Fig. \ref{scan}b. While the QBER is affected by many factors, the fringes observed in Fig. \ref{scan}a are likely due to the change in the phase relation between the master and the slave lasers \cite{Lau2009, Liu2020} as the detuning frequency and injection ratio are varied, which results in encoding errors in the relative phase between the slave laser pulses. From Fig. \ref{scan}b, the sparseness of the optimum regime can be clearly observed. The mapping of QBER takes more than 8 hours to complete, even when limiting to two parameters. This therefore highlights the need for an efficient method to determine the optimum operating regimes, especially in a large parameter space.

\begin{table}[ht]
\centering
\caption{Input parameters for the phase coherence and QBER optimisations.}
\begin{tabular}[t]{lcc}
\hline
&Phase coherence &QBER\\
\hline
1. Temperature of slave laser&\ding{51}&\ding{51}\\
2. DC bias of master laser&\ding{51}&\ding{51}\\
3. DC bias of slave laser&\ding{51}&\ding{51} \\
4. Injection power&\ding{51}&\ding{51} \\
5. Lasers temporal delay&\ding{51}&\ding{51}\\
6. RF modulation amplitude&&\ding{51}  \\

\hline
\label{parameters}
\end{tabular}
\end{table}%

\subsection{Phase coherence optimisation}

\begin{figure*}
    \centering
    \includegraphics{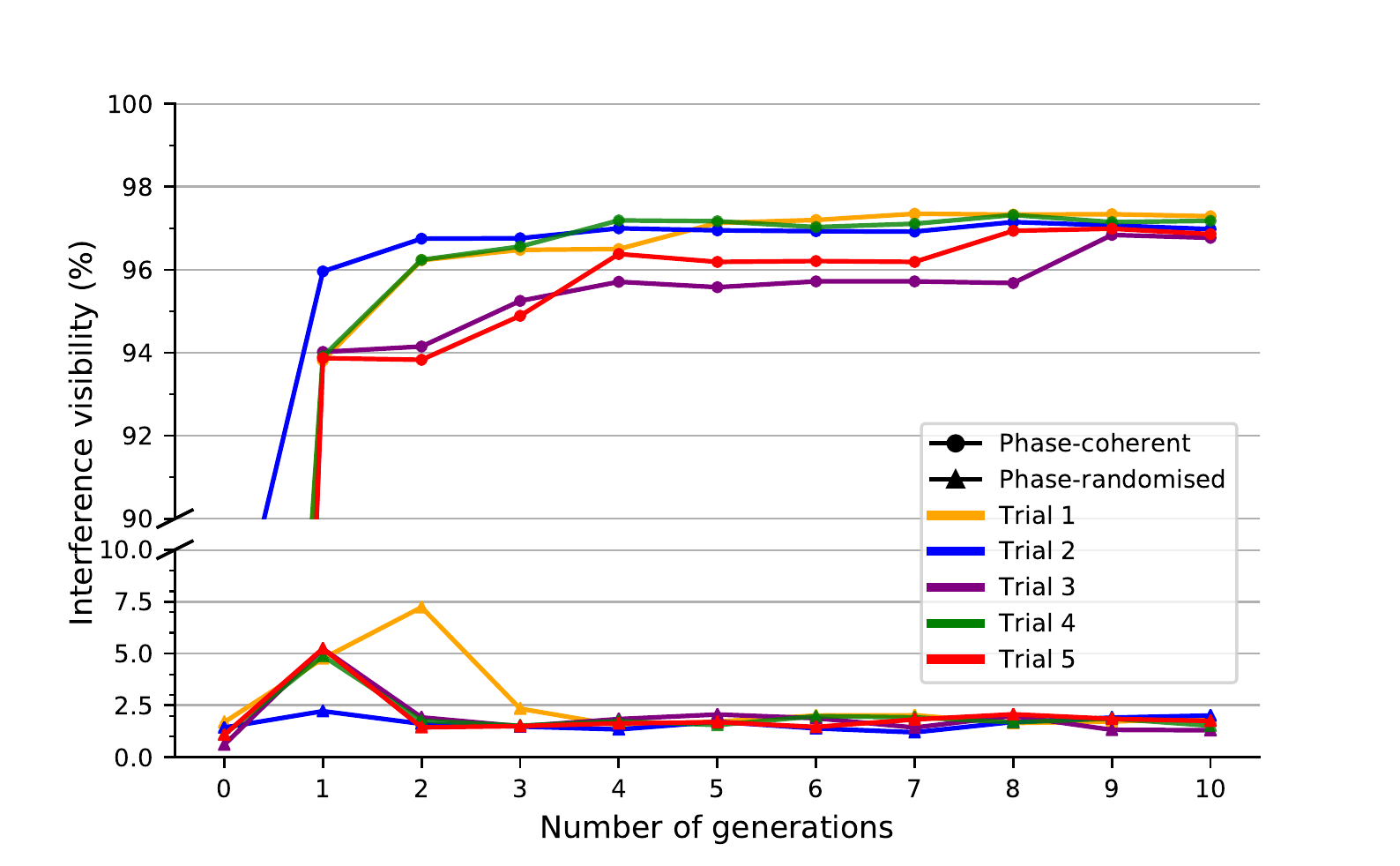}
    \caption{Evolution of optimisation on phase coherence for phase-coherent and phase-randomised pulses. The population size for the first three generations is 35, then reduced to 25 for subsequent generations. The time taken to complete 10 generations is 4 hours (50s per evaluation of individual).}
    \label{coherence_b}
\end{figure*}

Phase encoding is widely used in QKD protocols, where the secret bits are encoded in the relative phase between consecutive pulses \cite{Gisin2002}, thus, it is essential to have high phase coherence between pulses for deterministic phase control. In addition, a key requirement for secure quantum communication is that the phase of each qubit, comprised of the early and late time bins (Fig. \ref{waveform}), is uniformly random. This allows the coherent state of the attenuated pulses to be treated as photon number states and security proofs \cite{Gottesman2004, Lo2007} against the most general attacks can be obtained. 

OIL combined with gain switching represents a very efficient way to generate pulses that satisfy these requirements. As discussed in the previous section, gain switching allows each master pulse to carry a random phase while optical injection seeding allows the phase manipulation on the master pulse to be coherently transferred to the relative phase between consecutive slave laser pulses.

To investigate the phase coherence, the master laser is pulsed without additional modulation. As a result, the two slave laser pulses seeded by the same master pulse are in-phase, constructive and destructive interference can be obtained. In contrast, the slave laser pulses seeded by different master laser pulses have no definite phase relation, thus the interference should result in minimum visibility. To satisfy these conditions, we use the following fitness function which the algorithm aims to maximise by optimising the parameters shown in Table \ref{parameters}:

\begin{equation}
\psi_{coherence} = V_{coherent} + \frac{1}{V_{random}}
\end{equation}

\noindent where $V_{coherent}$ ($V_{random}$) is the interference visibility of the phase-coherent (phase-randomised) slave laser pulses.

The result of the optimisation is shown in Fig. \ref{coherence_b} where the performance of the best individual in the population over successive generations is plotted. The optimisation is initialised by assigning random values to the parameters from given ranges (i.e. within safe operating range). The probabilistic nature of the evolution and the random initial condition cause each optimisation to have a different trajectory. Thus, we repeat the optimisation for five times in order to capture all features as well as to verify its repeatability. As expected, through evolution, the algorithm learns to operate the system and the individuals in each new generation become increasingly competent as the quality of their genes improve. As the visibility for phase-coherent pulses is improved, the visibility of the phase-randomised pulses is also simultaneously suppressed over generations. Interestingly, while sometimes the visibility increases steadily over generations (trial 2 \& 4), it can also remain in local maxima for a few generations (trial 3 \& 5). Due to mutation and crossover, however, the algorithm eventually discovers a better operating regime, resulting in a sudden improvement after a plateau. As a result, all trials converge towards a visibility of $\sim$97\% for phase-coherent pulses and $<$2\% for phase-randomised pulses. The deviations from the theoretical ideal 100\% and 0\% visibilities are caused by the imperfections in real-world experimental equipment, e.g. imbalanced splitting ratio of the beam splitters in the interferometer, timing jitter of the pulses as well as system noise. The singularity in the fitness function is also avoided due to the non-zero minimum visibility in practice. Overall, the converged values
match with that obtained by tuning the transmitter manually. This shows that the algorithm is able to optimise the lasers to generate highly phase-coherent pulse pairs suitable for QKD encoding, while simultaneously ensuring each qubit has a globally random phase.

\subsection{QBER optimisation}

\begin{figure*}
    \begin{subfigure}
    \centering
    \includegraphics{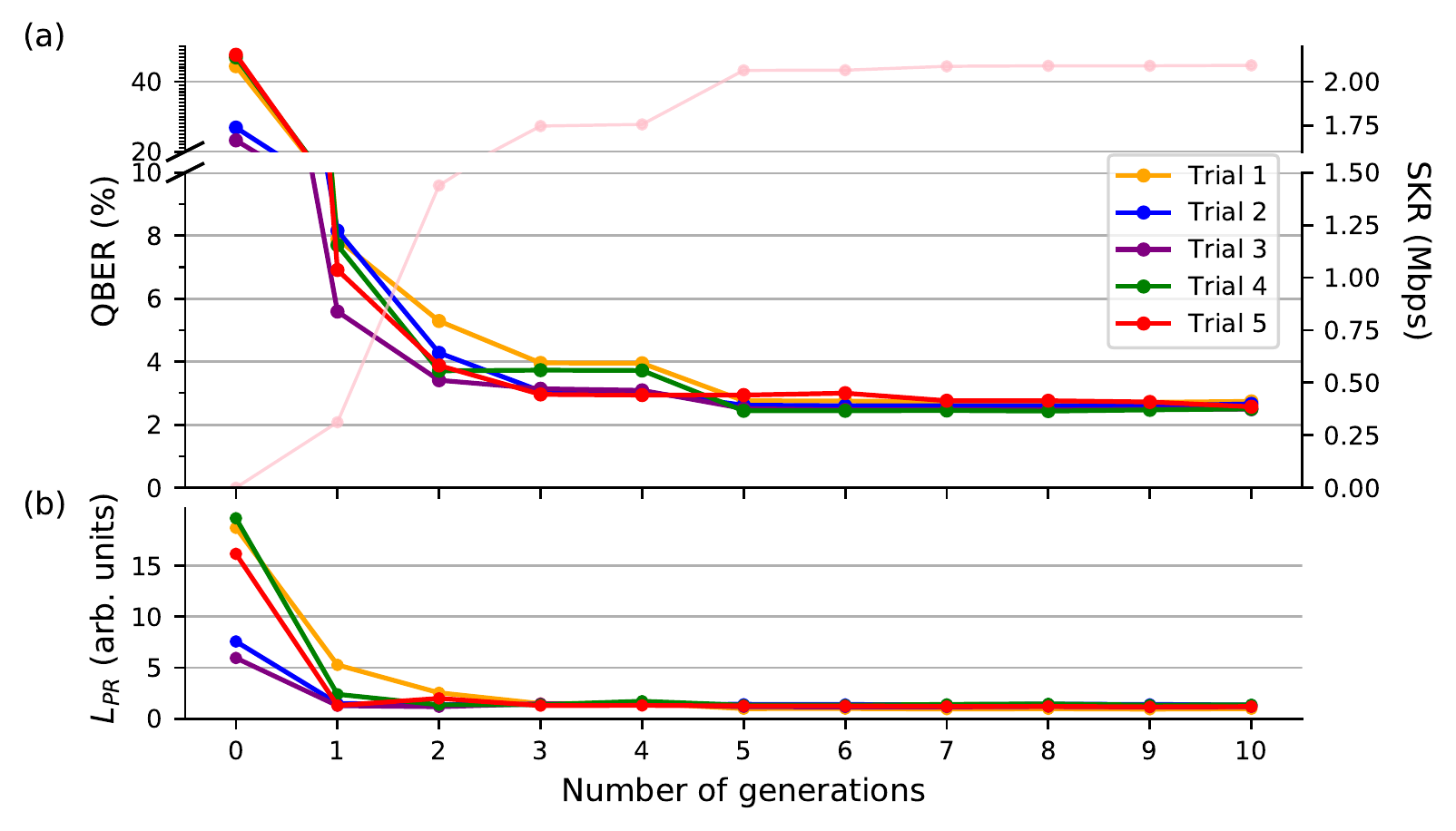}
    \end{subfigure}
    \begin{subfigure}
    \centering
    \includegraphics[width = 459pt]{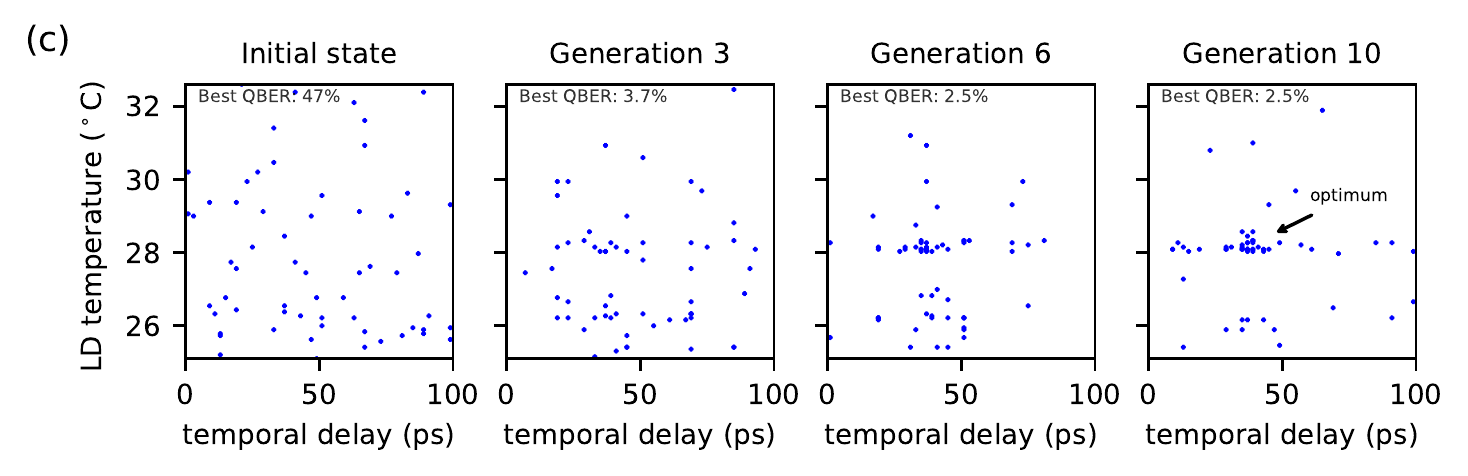}
     \end{subfigure}
        \caption{(a) Evolution of optimisation on (top) QBER, along with the corresponding secure key rate calculated based on the average QBER obtained over 5 trials and (b) loss function, $L_{PR}$. (c) Evolution of the population in the parameter space (only two dimensions are plotted, namely the slave laser temperature and the temporal delay between the master and slave lasers). The population size is 60 and the time taken to complete 10 generations is 2.5 hours (14s per evaluation of individual).}
        \label{qber_b}
\end{figure*}

The QBER is the primary measure of the performance of a practical QKD system. Minimising the QBER has been an indispensable task for QKD operations. Here we implement the aforementioned direct phase modulation scheme to encode random bits into the slave laser pulses (Fig. \ref{waveform}). After travelling through an optical channel with a loss of 16 dB (emulated by an VOA), the encoded pulses are decoded by the receiver AMZI and measured by the single photon detectors. We perform a proof-of-principle BB84 QKD protocol and optimise the QBER. 

As in phase coherence optimisation, it is important to take phase randomisation into account so that we can suppress the QBER while making sure that the global phase of the qubit is randomised at the same time. To achieve this, we exploit the fact that the intensity resulting from the interference between two phase-randomised pulses (referred to as side peak) is exactly half that of constructive interference between two phase-coherent pulses (referred to as signal peak), as illustrated in Fig. \ref{waveform}. Based on this, we define a phase randomisation `cost function', $L_{PR}$ which we aim to minimise:

\begin{equation}
 L_{PR} =  \alpha \, \frac{\abs{ \overline{C}_{signal} -  2\overline{C}_{side}}}{\overline{C}_{signal} } 
 \label{L_pr}
\end{equation}

\noindent where $\overline{C}_{signal}$ ($\overline{C}_{side}$) is the average photon counts measured from the signal peak (side peak) over an acquisition period. The scaling factor, $\alpha$ is chosen to be 1/10 in order to scale the $ L_{PR}$ so the contributions of phase randomisation and QBER are equally scored in the fitness function, which we define as:

\begin{equation}
\psi_{\mathrm{QBER}}=  \frac{1}{\mathrm{QBER}} + \frac{1}{L_{PR}}
\end{equation}

\begin{figure}
    \centering
    \includegraphics[width=246pt]{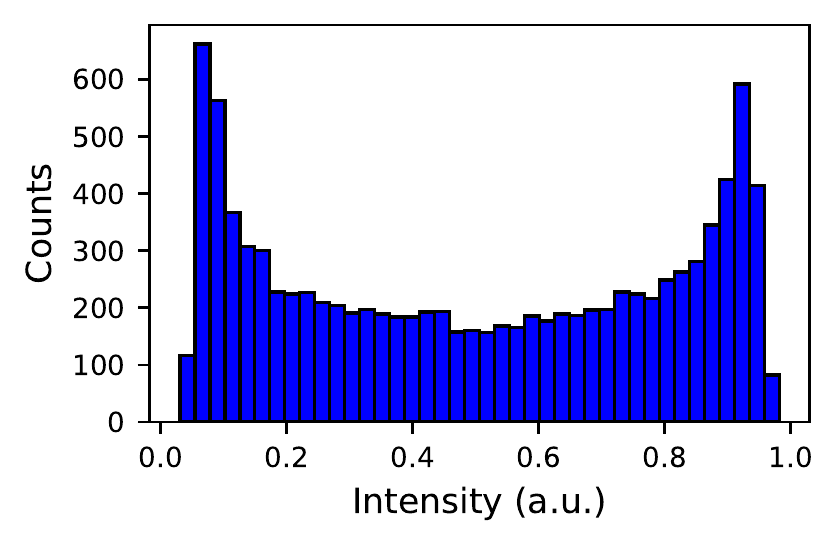}
    \caption{Intensity distribution from the interference between two consecutive pulses seeded by different master laser pulses, indicating random relative phase relation.}
    \label{phaserand}
\end{figure}

The input parameters are listed in Table \ref{parameters}. The evolution of the QBER and $L_{PR}$ of five repeated optimisation trials are plotted in Fig. \ref{qber_b}a-b. We perform a proof-of-principle QKD experiment using the BB84 protocol, where the secure key rate is estimated from gain and QBER measurements (Fig. \ref{qber_b}a). Similar to the phase coherence optimisation, all optimisation trials eventually locate the optimum parameters and converge towards a QBER of $\sim$2.5\%. Fig. \ref{qber_b}c further illustrates how the population evolves during the optimisation. At the beginning, the individuals are randomly distributed in the parameter space. After several generations, they gradually migrate towards the optimal region as the genes of the individuals from this region have a higher chance to be inherited by successive generations. Nevertheless, a small group of individuals remains scattered around the parameter space (through mutation) to keep exploring for even better operating regimes. To verify the phase randomisation, we removed the channel attenuation and measured the intensity probability distribution of the outputs with an oscilloscope, as shown in Fig. \ref{phaserand}, the distribution follows the typical profile expected from the interference between two phase-randomised pulses \cite{Yuan2014}.

\section{Discussion and conclusion}
In terms of the practical implementation, our GA-based optimisation technique can be seamlessly integrated into the software layer of QKD systems without requiring additional hardware modification. In particular, the QBER optimisation is designed to run within the QKD transmitter and receiver, without involving other diagnostic tools. Therefore, the optimisation procedure is self-contained and allows multiple QKD systems to be optimised automatically in parallel. This feature could be particularly useful in scaling up the manufacturing of the QKD systems, especially for chip-based QKD systems where the optimisation is often more challenging. For QKD systems deployed in real-world environments where testing equipment and QKD specialists are not readily available, our technique allows the QKD systems to be self-optimised in-situ in the case where the system parameters are detuned from optimum. Since information is disclosed between Alice and Bob to perform optimisation as part of the self-tuning process, we note that QKD keys are not generated during this time for security. Our GA-based optimisation technique is instead expected to be run to self-tune new systems and new in-field installations---QKD operation will commence once optimal parameters are found. Such optimal system parameters tend not to drift significantly, but the self-tuning algorithm could be re-run every few days or on-demand, to ensure long-term robust QKD system operation, even in the case of unexpected environment changes.

Regarding the performance of the optimisation, the speed of convergence depends on the complexity of the problem at hand as well as the control parameter configuration of the genetic algorithm. In order to efficiently locate the global optima, it is necessary to have sufficient gene diversity in the populations, especially the first generation so that the optima search does not overly rely on random mutations. Therefore, if the number of good solutions is very small compared to the size of the search space, a large population size is needed to maintain the diversity and avoid converging to local optima. However, having a larger population also inevitably increases the convergence times. It is well known that the control parameter configuration is problem-dependent. Here, the population size for phase coherence optimisation is chosen to be 35 and expanded to 60 for QBER optimisation due to its larger parameter space. These values are empirically determined to give repeatable convergence within reasonable amount of time and further optimisation is possible but beyond the scope of this work. 

Additionally, we note that our GA-based self-tuning technique is goal-oriented. Here we restrict our study to optical injection locking (OIL) in a QKD transmitter; yet it is straightforward for our approach to be applied more generally to other QKD protocols and system designs, e.g. QKD network or other optical communication light sources. Such an optimisation approach allows optimal operation to be achieved without a priori knowledge of the underlying complex dynamics.

In conclusion, we demonstrate a self-tuning QKD transmitter based on OIL by employing a GA. Through careful design of the algorithm configurations, we obtain consistent performance on the desired properties that match with that by manual optimisations, however, in a fully autonomous way. Our self-optimising approach therefore offers to make QKD systems and other quantum technologies more robust and practical.

\section{Data availability}
The datasets generated and analysed in the current study are available from the corresponding author upon reasonable request.\\

\begin{acknowledgments}
We acknowledge funding from the European Union Horizon 2020 Research and Innovation Programme (Grant 857156 “OPENQKD”). Y. S. L. acknowledges financial support from the EPSRC funded CDT in Delivering Quantum Technologies (EP/L015242/1) and Toshiba Europe Ltd. and V. L. acknowledges financial support from the EPSRC (EP/S513635/1) and Toshiba Europe Ltd.
\end{acknowledgments}


\section{Appendix}

\section*{QKD experiment}
We implement the two-decoy state BB84 protocol in the asymptotic case \cite{Ma2005}. The average photon numbers of the signal, decoy and vacuum states are 0.4, 0.1 and 0.001, respectively. The corresponding sending probabilities are 14/16, 1/16, 1/16 for signal, decoy and vacuum states, respectively. The X (Y) basis is chosen with a probability of 15/16 (1/16). The gain and QBER for each state are experimentally measured to calculate the final secure key rate.


\bibliography{library}



\end{document}